\documentstyle[preprint,aps]{revtex}
\begin{document}
\title{
\hspace{4.in} {\normalsize DOE/ER/40427-6-N96}\\
\hspace{4.in} {\normalsize ADP-96-24/T223}\\
\vspace{0.4cm}
Determination of flavor asymmetry for $\Sigma^{\pm}$\\
by the Drell-Yan process}

\author{Mary Alberg$^{a,b}$, Ernest M. Henley$^{b,c}$, Xiangdong
Ji$^{d}$\thanks{Present Address: Department of Physics and
Astronomy, University of Maryland, College Park, MD 20742} and
A.W. Thomas$^{e}$\\}

\address{{\it $^a$ 	Department of Physics,
			Seattle University,
			Seattle, WA 98122, USA}\\
	{\it $^b$ 	Department of Physics,
			Box 351560,
			University of Washington,
			Seattle, WA 98195-1560, USA}\\
	{\it $^c$ 	National Institute for Nuclear Theory,
			Box 351550,\\
			University of Washington,
			Seattle, WA 98195-1550, USA}\\
	{\it $^d$ 	Center for Theoretical Physics,
			Laboratory for Nuclear Science and
			Department of Physics,\\
			Massachusetts Institute of Technology,
			Cambridge, MA 02139, USA}\\
	{\it $^e$ 	Department of Physics and Mathematical Physics,\\
			and Institute for Theoretical Physics,
			University of Adelaide,
			Adelaide, S.A. 5005, Australia}}
\maketitle
\begin{abstract}
Flavor asymmetries for the valence and sea quarks of the
$\Sigma^{\pm}$ can be obtained from Drell-Yan experiments using charged
hyperon beams on proton and deuteron targets. A large, measurable
difference in sea quark asymmetries is predicted between SU(3) and 
pseudoscalar meson
models. The latter predict that in $\Sigma^{+}$, $\bar{u}/\bar{d}
\leq 1/2$, whereas the former predict $\bar{u}/\bar{d}
\approx 4/3$. Estimates of valence quark asymmetries based on quark
models also show large deviations from SU(3) predictions, which should
be measurable.
%
%
\vskip 5. in
\end{abstract}

\narrowtext
Parton distributions contain a wealth of information concerning the
non-perturbative structure of hadrons. While most attention
has so far been paid to the distributions of the nucleon, through the
Drell-Yan process one can also get information on the parton
distributions of the hyperons. We shall be concerned with the
predictions of the distibutions for the $\Sigma^{\pm}$ hyperons in
various models, including:
(1) a quark and meson model,
(2) an SU(3) model and (3) a model involving the coupling of octets of
mesons and baryons (without regard to mass). 
In particular, a comparison of $\Sigma^{\pm}$ and proton structure
functions can be used to differentiate between these models.
We show that measurements of the 
Drell-Yan process for $\Sigma^{\pm}$ scattering on
protons and neutrons (in deuteron targets) allow one to extract
information on the parton distributions of the $\Sigma^{\pm}$ and that
the expectations are quantitatively
quite different for (1), (2) and (3), above. In particular, we find
that measurements of the $\bar{u} /\bar{d}$ ratio in the $\Sigma^{+}$
are very sensitive to the model chosen, and that a quark-diquark
model of the hyperon predicts large SU(3) violations in 
the valence quark distributions.


One of the surprises in the structure of the proton is that the sea
appears to have a flavor asymmetry, an excess of $\bar{d}$ compared to
$\bar{u}$ \cite{NMC91,NMC94,NA51}. Although the experimental results
could imply some violation of isospin, this appears to be less likely,
and we interpret them as an $SU(2)_{Q}$ flavor asymmetry in the sea
 \cite{Forte93}. Thus, the $\bar{d}$ excess in the proton is expected to be
reflected in an excess of $\bar{u}$ in the neutron; isospin symmetry
would be broken if this were not the case. The evidence for flavor
asymmetry in the proton sea is based on analyses of deep inelastic 
muon scattering \cite{NMC91,NMC94} and Drell-Yan processes \cite{NA51}. One
explanation that has been offered is that the excess of $\bar{d}$
over $\bar{u}$ is due to the Pauli exclusion principle
\cite{Feynman77,Signal88,Signal89}. A more likely explanation, in our view,
is that offered by Thomas and colleagues
\cite{Thomas83,Ericson84,Thomas87,Melnitchouk91,Signal91}, Henley and
Miller \cite{Henley90}, and others
\cite{Eichten92,Kumano91a,Kumano91b,Hwang91,Szczurek93,Szczurek94,Szczurek96,%
Holtmann96},
namely that the presence of a
pion cloud surrounding a proton favors $\bar{d}$ over $\bar{u}$
because of the excess positive charge of the meson cloud.

It is interesting to apply these arguments to the strange
baryons and to compare them with SU(3). Here we focus on the charged
$\Sigma^{+}$ ($\Sigma^{-}$), composed
of $uus$ ($dds$) valence quarks. Thus the main difference from the
$p(n)$ case is the replacement of a valence
$d(u)$ quark by an $s$ quark. In the following, the quark distribution
 $q(u,d,$ or $s)$
without subscripts refers to the proton, and with a $\Sigma$ subscript
refers to the $\Sigma^{+}$. The $x$-dependence of these
distributions is implied, but not shown explicitly. With neglect of
mass 
effects or under SU(3), 

\begin{equation}
\bar{r}\equiv \frac{\bar{u}}{\bar{d}} =\frac{\bar{u}_{\Sigma}}
{\bar{s}_{\Sigma}}=0.51\pm0.04\pm0.05 ,
\end{equation}
where the experimental ratio \cite{NA51} is that obtained for the
proton at $x\approx
0.18$.

The ratio $\kappa \equiv 2 \bar{s}/(\bar{d}+\bar{u})$ is a measure of
the strange quark content of the nucleon. It has been determined
experimentally in neutrino-induced charm production
\cite{CDHS,CCFR,E733} to be in the range
$0.373^{+0.048}_{-0.041}\pm{0.018}\leq \kappa \leq 0.57 \pm 0.09$.
The CTEQ \cite{CTEQ} determination of parton
distributions from global QCD analyses 
of experimental data uses the value $\kappa = 0.5$, from which
\begin{equation}
\bar{r}_{s}\equiv \frac{\bar{s}}{\bar{u}+\bar{d}}=\frac{\kappa}{2}=0.25,
\end{equation}
so with $\bar{d}\approx 2\bar{u}$ from Eq. 1,
\begin{equation}
\frac{1}{\bar{r}_{\Sigma}}\equiv
\frac{\bar{d}_{\Sigma}}{\bar{u}_{\Sigma}}=\frac{\bar{s}}{\bar{u}}\approx 0.75 
 .
\end{equation}
We also find
\begin{equation}
\frac{\bar{d}_{\Sigma}}{\bar{s}_{\Sigma}}=\frac{\bar{s}}{\bar{d}}\approx 0.38
\end{equation}
from the same analysis.

For the model of quarks surrounded by light pseudoscalar mesons, 
the $\Sigma^{+}$
has an excess of $\bar{d}$ over $\bar{u}$ (and the opposite for the
$\Sigma^{-}$) in contradistinction to the SU(3) prediction. If we
neglect higher masses, then the $\Sigma^{+}(uus)$ will have components
$\Lambda^{0}(uds)\pi^{+}(u\bar{d})$,
$\Sigma^{0}(uds)\pi^{+}(u\bar{d})$,
$\Sigma^{+}(uus)\pi^{0}(\frac{1}{\sqrt{2}}[d\bar{d}-u\bar{u}])$, or
$p(uud)\bar{K^{0}}(\bar{d}s)$; similarly a $\Sigma^{-}(dds)$ can be
$\Lambda^{0}(uds)\pi^{-}(d\bar{u})$,
$\Sigma^{0}(uds)\pi^{-}(d\bar{u})$,
$\Sigma^{-}(dds)\pi^{0}(\frac{1}{\sqrt{2}}[d\bar{d}-u\bar{u}])$, or
$n(udd)K^{-}(\bar{u}s)$. Thus there is a clear enhancement of
$\bar{d}$ for $\Sigma^{+}$ and $\bar{u}$ for $\Sigma^{-}$ and we expect $
{\bar{r}_{\Sigma}} \leq$ 0.5.

We also consider an SU(3) model in which a baryon is composed of 
octets of baryons and mesons.
We use the SU(3) isoscalar factors and representation matrices given by the
Particle Data Group\cite{PDG}. 
For \mbox{\boldmath $8_{1}\rightarrow 8 \otimes 8$}
\begin{equation}
N \rightarrow \frac{g_{1}}{\sqrt{20}}\left[3N\pi-N\eta-3\Sigma
K-\Lambda K\right],
\end{equation}
and for \mbox{\boldmath $8_{2}\rightarrow 8 \otimes 8$}
\begin{equation}
N \rightarrow \frac{g_{2}}{\sqrt{12}}\left[\sqrt{3}N\pi+\sqrt{3}N\eta+
\sqrt{3}\Sigma K-\sqrt{3}\Lambda K\right].
\end{equation}
The standard $D$ and $F$ couplings are
related to $g_{1}$ and $g_{2}$ by $D=\frac{\sqrt{30}}{40}g_{1}$ and
$F=\frac{\sqrt{6}}{24}g_{2}$, so
\begin{eqnarray}
p& \rightarrow &\sqrt{8}D\left[p\left\{(1+\alpha)\pi^{0}+
\sqrt{3}\left(\alpha-\frac{1}{3}\right)\eta\right\}
-\sqrt{2}(1+\alpha)n\pi^{+}+\sqrt{2}(\alpha-1)\Sigma^{+}K^{0}\right.%
\nonumber \\
&
&\left.+(1-\alpha)\Sigma^{0}K^{+}-\sqrt{3}\left(\alpha+\frac{1}{3}\right)%
\Lambda K^{+}\right],
\end{eqnarray}
\begin{eqnarray}
p& \rightarrow &
\sqrt{8}D\left[p\left\{(1+\alpha)\frac{u\bar{u}-d\bar{d}}
{\sqrt{2}}+\sqrt{3}\left(\alpha-\frac{1}{3}\right)
\frac{u\bar{u}+d\bar{d}-2s\bar{s}}{\sqrt{6}}\right\}-\sqrt{2}
(1+\alpha)nu\bar{d}\right.\nonumber
\\
& & \left.+
\sqrt{2}(\alpha-1)\Sigma^{+}d\bar{s}+(1-\alpha)\Sigma^{0}u\bar{s}-
\sqrt{3}\left(\alpha+\frac{1}{3}\right)\Lambda u\bar{s}\right],
\end{eqnarray}
with $\alpha \equiv F/D$, and which leads to relative probabilities, 
averaged over $x$,
\begin{eqnarray}
\bar{u} & \approx & \frac{2}{9}\left[9\alpha^{2}+6\alpha+1\right],\\
\bar{d} & \approx & \frac{2}{9}\left[9\alpha^{2}+18\alpha+13\right],\\
\bar{s} & \approx & \frac{4}{9}\left[18\alpha^{2}-12\alpha+8\right].
\end{eqnarray}
Then
\begin{equation}
\bar{r}\equiv \frac{\bar{u}}{\bar{d}}=\frac{\bar{u}_{\Sigma}}
{\bar{s}_{\Sigma}}=
\frac{1+6F/D+9(F/D)^2}{13+18F/D+9(F/D)^2}
\end{equation}
and
\begin{equation}
\bar{r}_{s}\equiv
\frac{\bar{s}}{\bar{u}+\bar{d}}=\frac{\bar{d}_{\Sigma}}
{\bar{u}_{\Sigma}+\bar{s}_{\Sigma}}=\frac{8-12F/D+18(F/D)^2}{7+12F/D+9(F/D)^2}.
\end{equation}
With $\alpha = 0.6$, consistent with a recent analysis\cite{Ratcliffe}, 
we obtain for the proton
\begin{equation}
\bar{r}=.29,\: \bar{r}_{s}=.42
\end{equation}
which differ significantly from the experimental result
($\bar{r}=.51$) and parameter ($\bar{r}_{s}= \kappa /2 = .25$). We also show in
Table I the prediction of this model, $\bar{r}_{\Sigma}=0.54$,
which, like the meson cloud model, disagrees with the SU(3)
expectation of 4/3.

Deviations from SU(3) predictions are also expected for the valence
quark distributions in $\Sigma^{+}$ ($\Sigma^{-}$). On the basis of
SU(3) symmetry we expect

\begin{equation}
r_{\Sigma} \equiv \frac {s_{\Sigma}}{u_{\Sigma}} \approx \frac {d}{u} 
\approx 0.57 (1-x).
\label{SU3}
\end{equation}
The functional form is taken from a fit by CDHS \cite{CDHSV}, and
agrees with the latest parton distribution analysis of CTEQ
\cite{CTEQ} within $20\%$, which is adequate for our calculations.

We find that for $x \geq 0.2$, quark models predict valence quark
flavor asymmetries
in the $\Sigma^{+}$ that are greater than the SU(3) result, e.g. by a
factor of 3.4 at $x=0.7$. Our approach to estimating valence quark
distributions in the $\Sigma^{+}$ is based on a quark-diquark model 
initiated at
Adelaide \cite{Signal89,Schreiber90} which has led to the
study of charge symmetry
violation in the
nucleon \cite{CSV}. 
It was found that the dominant contribution to the structure
function $q(x)$ in the valence region comes from a state in which the two
spectator quarks are in their ground states. The effective mass of this diquark
state will deviate from 3/4 of the nucleon mass (in the MIT bag model,
2/3 in the constituent quark model) because of the
hyperfine interaction. The mass difference between the two spin states
of the diquark leads to spin and flavor dependence of $q(x)$ \cite{Close88}.
Let $x_{q}$ represent the most probable momentum fraction carried by
the quark $q$,
and $x_{qq}$ represent the most probable momentum fraction carried by the
diquark. Then the peak in $q(x)$ can be estimated from
\begin{equation}
x_{q}+x_{qq}=1,\: x_{q}=1-x_{qq}\approx 1-\frac{m_{qq}}{m_{B}},
\end{equation}
in which $m_{qq}$ and $m_{B}$ are the diquark and baryon masses, respectively.
For the nucleon, the $N-\Delta$ splitting leads to $m_{qq}=650$ MeV in
the spin singlet state, and $m_{qq}=850$ MeV in the spin triplet. Then
in the proton, $d(x)$ peaks at $x_{d} \approx 0.10$, whereas $u(x)$
peaks at $x_{u} \approx 0.31$ -- at the scale appropriate to the model. 
After QCD evolution, these estimates are in reasonable
agreement with recent parton distribution analyses \cite{CTEQ}.

These same arguments may be applied to the $\Sigma^{+}$. In this case
the diquark $uu$ must be in a spin triplet, so from the
$\Lambda-\Sigma$ splitting, $m_{uu} \approx 850$ MeV, and $s_{\Sigma}(x)$ peaks
at $x_{s} \approx 0.28$. This is close to the value found for $u(x)$
in the proton, so we set $s_{\Sigma}(x)\approx u(x)/2$ (the factor of
2 comes from normalization). To estimate the $u_{\Sigma}$ distribution
we note that the $su$ diquark
mass is increased by $\approx 180$ MeV because of $m_{s}$, and with
the hyperfine splitting,  $m_{su} \approx 900$ MeV in the
singlet, leading to a peak $x_{u}(S=0) \approx 0.24$ -- i.e., a ``harder''
distribution, like that of $u(x)$ in the proton -- 
and $m_{su} \approx 1050$ MeV in
the triplet, leading to $x_{u}(S=1) \approx 0.10$ -- 
a ``softer'' distribution like $d(x)$ in
the proton. Since the singlet and triplet diquark states are equally
probable, we approximate $u_{\Sigma}(x) \approx d(x)+u(x)/2$. Then
\begin{equation}
r_{\Sigma} \equiv \frac{s_{\Sigma}}{u_{\Sigma}}\approx \frac
{u}{2(d+u/2)} \approx
\frac{1}{1+2d/u} =\frac{1}{1+1.14(1-x)}.
\label{QM}
\end{equation}
In the valence quark region, this ratio is considerably in
excess of that predicted by SU(3), as can be seen in Fig. 1, in which
we plot the ratio $R$
\begin{equation}
R=\frac{(r_{\Sigma})_{th}}{(r_{\Sigma})_{SU(3)}}
\approx \frac{1}{[1+1.14(1-x)]0.57(1-x)}
\end{equation}

There are a number of ways that these arguments and models can be
tested for $\Sigma$ hyperons. The most practical to us appears to be
 in terms of the Drell-Yan
cross sections for $\Sigma^{\pm}p$ and $\Sigma^{\pm}n$ (i.e. $d$) -- 
e.g., in the inclusive reactions $\Sigma^{\pm}p\rightarrow l^{+} l^{-}
X$, where $l^{\pm}$ are muons or electrons and $X$ is
unmeasured. Beams of $\Sigma^{\pm}$ appear to be adequate for this
purpose, but $\pi$ contamination will lead to problems which need to be
overcome. 

We first consider the determination of sea quark flavor asymmetry for
$\Sigma^{\pm}$. We find that extracting the ratio $\bar{r}_{\Sigma}(x)\equiv
\bar{u}_{\Sigma}(x)/\bar{d}_{\Sigma}(x)$ for the $\Sigma^{+}$ depends on
the known ratios $r(x)\equiv u(x)/d(x)$ and
  $\bar{r}(x)\equiv \bar{u}/\bar{d}$ in the
proton. The former is well-determined, and the recent determination of the
latter has been discussed above. Ratios involving
$\bar{s}$ in the $\Sigma^{\pm}$ cannot be tested easily
because they involve second order annihilations ($\bar{s}_{\Sigma}$ on
$s$), terms which we neglect because the present 
accuracy of
Drell-Yan measurements is insufficient to be sensitive to them. 

Drell-Yan cross-sections are proportional to the products
$q(x)\bar{q}(x^{\prime})$, weighted by the product of the quark charges, and
summed over contributions from beam and
target. We neglect sea-quark - sea-quark collisions, which would
contribute below the likely level of accuracy of the experiment. 
We assume isospin reflection (charge) symmetry: $u(x)=d_{n}(x)$,
$\bar{u}(x)=\bar{d}_{n}(x)$, $u_{\Sigma^{+}}(x)=d_{\Sigma^{-}}(x)$,
$\bar{u}_{\Sigma^{+}}(x)=\bar{d}_{\Sigma^{-}}(x)$, and
$s_{\Sigma^{+}}(x)
=s_{\Sigma^{-}}(x)$.
 In
the following equations, $q(x)$ represents valence quarks and
$\bar{q}(x)$ represents sea quarks. The valence quark
normalizations are: $\int u(x)\,dx = 2$ and $\int d(x)\,dx = 1$.

Consider the Drell-Yan process for $\Sigma N$. Let $\sigma(\Sigma N)$
represent the cross-section for inclusive dilepton production
\begin{equation}
\sigma(\Sigma N) \equiv s\frac{d^{2}\sigma (\Sigma N \rightarrow
l^{+}l^{-}X)}{d\sqrt{\tau}dy}=\frac{8\pi\alpha^2}{9\sqrt{\tau}}
K(x_{\Sigma},x_{N})\sum_{i}e_{i}^{2}\{q_{i}(x_{\Sigma})\bar{q}_{i}(x_{N})
+[\Sigma \leftrightarrow N]
\label{ddcross}
\end{equation}
with $M$ the mass of the dilepton pair and
$\sqrt{\tau}=M/\sqrt{s}$. The factor $K(x_{\Sigma},x_{N})$ accounts
for higher-order
QCD corrections. If the c.m. rapidity $y \approx 0$, then 
$x_{\Sigma}\approx x_{N}\approx x$, and
\begin{equation}
\sigma (\Sigma^{+}p)\approx \frac{8\pi\alpha^2}{9\sqrt{\tau}}K(x)
\{ \frac{4}{9}[u(x)\bar{u}_{\Sigma}(x)+u_{\Sigma}(x)\bar{u}(x)]+
\frac{1}{9}[d(x)\bar{d}_{\Sigma}(x)+s_{\Sigma}(x)\bar{s}(x)]\}.
\label{cross1}
\end{equation}
Then by charge symmetry
\begin{equation}
\sigma (\Sigma^{-}n)\approx \frac{8\pi\alpha^2}{9\sqrt{\tau}}K(x)
\{ \frac{1}{9}[u(x)\bar{u}_{\Sigma}(x)+u_{\Sigma}(x)\bar{u}(x)+
s_{\Sigma}(x)\bar{s}(x)]+\frac{4}{9}d(x)\bar{d}_{\Sigma}(x)\}.
\end{equation}
We also find
\begin{equation}
\sigma (\Sigma^{+}n)\approx \frac{8\pi\alpha^2}{9\sqrt{\tau}}K(x)
\{\frac{4}{9}[d(x)\bar{u}_{\Sigma}(x)+u_{\Sigma}(x)\bar{d}(x)]+
\frac{1}{9}[u(x)\bar{d}_{\Sigma}(x)+s_{\Sigma}(x)\bar{s}(x)]\},
\end{equation}
and again by charge symmetry
\begin{equation}
\sigma (\Sigma^{-}p)\approx \frac{8\pi\alpha^2}{9\sqrt{\tau}}K(x)
\{\frac{1}{9}[d(x)\bar{u}_{\Sigma}(x)+u_{\Sigma}(x)\bar{d}(x)+
s_{\Sigma}(x)\bar{s}(x)]+\frac{4}{9}u(x)\bar{d}_{\Sigma}(x)\}.
\label{cross4}
\end{equation}
 As we note below, if $K(x)$ is known, and all four cross sections are
 measured,
$\bar{u}_{\Sigma}$, $\bar{d}_{\Sigma}$, ${u}_{\Sigma}$, and
${s}_{\Sigma}$ can be determined. The uncertainties in
$K(x)$ can be
factored out by taking ratios of cross
sections; two independent ratios can be constructed. We first define a ratio
$R^{\prime}(x)$ determined from
the Drell-Yan cross-sections so as to eliminate all unknowns except
for $\bar{r}_{\Sigma}(x)$
\begin{equation}
R^{\prime}(x)\equiv \frac {[\sigma (\Sigma^{+}p)-\sigma
(\Sigma^{-}n)]+\bar{r}(x)[\sigma (\Sigma^{-}p)-\sigma (\Sigma^{+}n]}
	      {[\sigma (\Sigma^{+}p)-\sigma (\Sigma^{+}n)]+4[\sigma 
(\Sigma^{-}p)-\sigma (\Sigma^{-}n)]},
\end{equation}
and use Eq. \ref{cross1} - Eq. \ref{cross4} to write $R^{\prime}(x)$
in terms of the ratios
$\bar{r}_{\Sigma}(x)$, $r(x)$ and $\bar{r}(x)$:

\begin{equation}
R'(x)=\frac{\bar{r}_{\Sigma}(x)[r(x)-\bar{r}(x)]-[1-\bar{r}(x)r(x)]}
       {5[r(x)-1]}.
\label{R'}
\end{equation}
Thus for $r(x)\approx 2$ and $\bar{r}(x)\approx 0.5$,
$R{^\prime}(x)\approx 0.3 \,\bar{r}_{\Sigma}(x)$.

If $K(x)$ is known, $\bar{d}_{\Sigma}(x)$ can be determined directly
from the cross sections:

\begin{equation}
\bar{d}_{\Sigma}(x)=\frac{27\sqrt{\tau}}{40\pi\alpha^2 K(x)}\frac
{[\sigma (\Sigma^{+}p)-\sigma (\Sigma^{+}n)]+4[\sigma (\Sigma^{-}p)-
\sigma (\Sigma^{-}n)]}{[u(x)-d(x)]},
\end{equation}
and $s_{\Sigma}(x)$ can be determined from the cross sections and $\bar{s}(x)$:

\begin{equation}
s_{\Sigma}(x)=\frac{27\sqrt{\tau}}{8\pi\alpha^2 K(x)}\frac
{[\sigma (\Sigma^{+}n)-4\sigma (\Sigma^{-}p)]-r(x)[\sigma 
(\Sigma^{+}p)-4\sigma (\Sigma^{-}n)]}{\bar{s}(x)[r(x)-1]}.
\end{equation}
(Recall that, because of the higher mass of the strange quark, we expect
$s_{\Sigma}(x)$ to peak at a larger $x$ than $d(x)$ --c.f., Eq. 11.)

Quark models with a meson cloud predict the sea quark distributions
$\bar{q}(x)$; they also predict that the difference $D\equiv x
[\bar{d}(x)-\bar{u}(x)]$ peaks at $x\approx 0.1$
\cite{Eichten92,Kumano91a,Kumano91b,Hwang91,Szczurek93,Szczurek94,Szczurek96}.
On the basis of meson cloud models\footnote{We are undertaking a calculation
of the $\Sigma^{\pm}$ sea quark distributions.}, the distributions of sea
quarks in
the $\Sigma^{\pm}$ may differ somewhat from those in the nucleon due to
the presence of kaons; this may shift the maximum of $D$ to somewhat
smaller values of $x$. Nevertheless, the region $0\leq x \leq 0.2$
should be a good one in which to determine $\bar{r}_{\Sigma}$.

We believe that the measurement of $R^{\prime}$ should be possible to
 within $\approx
20\%$ and this is sufficient to establish the preponderance of
$\bar{d}$ over $\bar{u}$ in the $\Sigma^{+}$, as predicted by the
octet and meson cloud models. From Eq. \ref{R'}, an error, $e$, in the
measurement of $R^{\prime}$ leads to an error of
approximately $3e$ in $\bar{r}_{\Sigma}$. If $\bar{r}_{\Sigma}$ were
found to be $\leq 0.5$, together with the known
value of $\bar{r}\approx 0.51$, this measurement would help to
reinforce the necessity to include pseudoscalar mesons in 
quark models of baryons.

%
%
To measure valence quark asymmetries we consider the Drell-Yan
process for $\Sigma^+$ and $\Sigma^-$ on isoscalar targets -- with cross
sections $\sigma_+$ and $\sigma_-$, respectively. We fix $x_{\Sigma}$
 to be above 0.3 so that valence
quarks in the hyperons dominate. Then from Eq. \ref{ddcross}-
Eq. \ref{cross4}, with $x\equiv
x_{\Sigma}$ and $x'\equiv x_{N}$,
\begin{equation}
\sigma_{+}\equiv \sigma (\Sigma^{+}A)\approx
\frac{8\pi\alpha^2}{9\sqrt{\tau}}\frac{A}{2}
K(x,x')
\{ \frac{4}{9}u_{\Sigma}(x)[\bar{u}(x')+\bar{d}(x')]+\frac{2}{9}
[s_{\Sigma}(x)\bar{s}(x')]\},
\end{equation}
and
\begin{equation}
\sigma_{-}\equiv \sigma (\Sigma^{-}A)\approx
\frac{8\pi\alpha^2}{9\sqrt{\tau}}\frac{A}{2}
K(x,x')
\{ \frac{1}{9}u_{\Sigma}(x)[\bar{u}(x')+\bar{d}(x')]
+\frac{2}{9}[s_{\Sigma}(x)\bar{s}(x')]\}.
\end{equation}
We approximate $K(x,x')$ by an average $\bar{K}$, and integrate over
 $x^{\prime}$ in
the nucleons, so that 
\begin{equation}
\int dx^{\prime}\,\sigma_{+}(x^{\prime},x)=
\frac{8\pi\alpha^2}{9\sqrt{\tau}}\frac{A}{2}
\bar{K}
\{ \frac{4}{9}u_{\Sigma}(x)[\bar{u}+\bar{d}]+
\frac{2}{9}[s_{\Sigma}(x)\bar{s}]\},
\end{equation}
and similarly for $\sigma_{-}$, with
$\bar{q} = \int dx^{\prime}\,\bar{q}(x^{\prime})$. Then
\begin{equation}
R_{v}(x)\equiv \frac{\int dx^{\prime}\,\sigma_{-}(x^{\prime},x)}
		{\int dx^{\prime}\,\sigma_{+}(x^{\prime},x)}
	=\frac{u_{\Sigma}(x)(\bar{d}+\bar{u})+2s_{\Sigma}(x)\bar{s}}
	     {4u_{\Sigma}(x)(\bar{d}+\bar{u})+2s_{\Sigma}(x)\bar{s}}
	=\frac{1+\kappa r_{\Sigma}}{4+\kappa r_{\Sigma}}.
\label{exp}
\end{equation}
We again use the CTEQ \cite{CTEQ} value, $\kappa=0.5$, and evaluate
$R_{v}$
 for both the SU(3) prediction for $r_{\Sigma}$
(Eq. \ref{SU3}) and for our quark model (Eq. \ref{QM}). In Fig. 2 we plot 
the ratio $D$
\begin{equation}
D(x) \equiv \frac{R_{v}{\rm (quark \: model)}}{R_{v}{\rm (SU(3))}}.
\end{equation}
We note that the predicted asymmetry 
exceeds the SU(3) prediction by about $10\%$
at $x=0.5$, increasing to $20\%$ at $x=0.75$. An accuracy of $\approx
5\%$ should be possible for these integrated cross sections. Thus
these measurements will test the SU(3) violations in the valence quark
distributions of $\Sigma^{\pm}$ predicted by quark models.
 
In summary, there are substantial differences expected between the
valence and sea parton distributions associated with several models of
hyperon structure. We have seen that Drell-Yan experiments based on
existing hyperon beams should be capable of testing these ideas. The
substantial violations of SU(3) flavor symmetry in the valence
distributions are probably the easiest to test as they require only an
isoscalar target and a semi-integrated cross-section. However, the
enormous interest in the underlying cause of the flavor asymmetry of the
proton sea should also make the tests of sea quark distributions an
important priority as well.

This work has been supported in part by the U.S. Department of Energy,
Contract \# DOE/ER/4027-6-N96, by the National Institute for Nuclear
Theory and by the Australian Research Council. 
We wish to thank Joel Moss, Jen-chieh Peng and other
participants in the program INT-96-1, ``Quark and Gluon Structure of
Nucleons and Nuclei'' for helpful discussions, and Jen-chieh Peng for a 
constructive critique of this manuscript. XJ and AWT also thank the 
Institute for Nuclear Theory for its hospitality during the time that part 
of this work was done.

\begin{table}
\caption{Sea quark asymmetries}
\vspace{0.4cm}
\begin{center}
\begin{tabular}{ccccc}
	& meson cloud & SU(3) & octets & experiment \\ 
\hline
$\bar{r} \equiv \frac{\bar{u}}{\bar{d}}$ &
theory ref &
 &
0.29 &
0.51  \\
$\bar{r}_{s} \equiv \frac{\bar{s}}{\bar{u}+\bar{d}}$ &
&
&
0.42 &
$\frac{\kappa}{2}=0.25$\\  
$\bar{r}_{\Sigma} \equiv \frac{\bar{u}_{\Sigma}}{\bar{d}_{\Sigma}}$ &
$\leq 0.5$ &
4/3 &
0.54 &

\end{tabular}
\end{center}
\end{table}

\begin{figure}
\caption{Plot of the ratio $R$, Eq. 18, as a function of $x$. 
We also show 
$r_{\Sigma}$, as obtained from the quark model, Eq. 17, and as found from 
SU(3), Eq. 15.}
\end{figure}
\par
\begin{figure}
\caption{Plot of D, Eq. 32, as a function of $x$.}
\end {figure}

\begin{references}
\bibitem{NMC91} P. Amaudruz et al., Phys. Rev. Lett. 66 (1991) 2712.
\bibitem{NMC94} M. Arneodo et al., Phys. Rev. D 50 (1994) R1.
\bibitem{NA51} A. Baldit et al., Phys. Lett. B 332 (1994) 244. 
\bibitem{Forte93} S. Forte, Phys. Rev. D 47 (1993) 1842.
\bibitem{Feynman77} R.P. Feynman and R.D. Field, Phys. Rev. D 15 (1977) 2590.
\bibitem{Signal88} A.I. Signal and A.W. Thomas, Phys. Lett. B 221
(1988) 481.
\bibitem{Signal89} A.I. Signal and A.W. Thomas, Phys. Rev. D 40 (1989) 2832.
\bibitem{Thomas83} A.W. Thomas, Phys. Lett. 126B (1983) 97.
\bibitem{Ericson84} M. Ericson and A.W. Thomas, Phys. Lett. 148B (1984) 191.
\bibitem{Thomas87} A.W. Thomas, Prog. Theor. Phys. [Suppl.] 91 (1987) 204.
\bibitem{Melnitchouk91} W. Melnitchouk, A.W. Thomas, and A.I. Signal,
Z. Phys. A 340 (1991) 85.
\bibitem{Signal91} A.I. Signal, A.W. Schreiber, and A.W. Thomas,
Mod. Phys. Lett. A 6 (1991) 271.
\bibitem{Henley90} E.M. Henley and G.A. Miller, Phys. Lett. B 251 (1990) 453.
\bibitem{Eichten92} E. Eichten, I. Hinchliffe, and C. Quigg,
Phys. Rev. D 45 (1992) 2269.
\bibitem{Kumano91a} S. Kumano, Phys. Rev. D 43 (1991) 59.
\bibitem{Kumano91b} S. Kumano and J.T. Londergan, Phys. Rev. D 44 (1991)
717.
\bibitem{Hwang91} W-Y. P. Hwang, J. Speth, and G.E. Brown, Z. Phys. A
339 (1991) 383.
\bibitem{Szczurek93} A. Szczurek and J. Speth, Nucl. Phys. A555 (1993)
249.
\bibitem{Szczurek94} A. Szczurek, J. Speth, and G.T. Garvey,
Nucl. Phys. A570 (1994) 765.
\bibitem{Szczurek96} A. Szczurek, M. Ericson, H. Holtmann, and
J. Speth, Nucl. Phys. A596 (1996) 397.
\bibitem{Holtmann96} H. Holtmann, N.N. Nikolaev, J. Speth, and
A. Szczurek, Z. Phys. A 353 (1996) 411.
%
\bibitem{CDHS} H. Abromowicz et al., Z. Phys. C 15 (1982) 19.
\bibitem{CCFR} C. Foudas et al., Phys. Rev. Lett. 64 (1990) 1207;
M.H. Shaevitz, Nucl. Phys. B (Proc. Suppl.) 19 (1991) 270;
S.A. Rabinowitz et al., Phys. Rev. Lett. 70 (1993) 134; A.O. Bazarko
et al., Z. Phys. C 65 (1995) 189.
\bibitem{E733} B. Strongin et al., Phys. Rev. D 43 (1991) 2778.
\bibitem{CTEQ} H.L. Lai et al.,
Phys. Rev. D 51 (1995) 4763; CTEQ -604, hep-ph/9606399 (1996).
\bibitem{PDG} R. M. Barnett et al., Phys. Rev. D54 (1996) 173.
\bibitem{Ratcliffe} P. G. Ratcliffe, Phys. Lett. B365 (1996) 383.
\bibitem{CDHSV} F. Eisele, J. de Physique C3 (Suppl) (1982) C3.
%
\bibitem{Schreiber90} A.W. Schreiber, A.W. Thomas, and J.T. Londergan,
Phys. Rev. D 42 (1990) 2226.
%
\bibitem{CSV}E. Sather, Phys. Lett. B274 (1992) 433; \\
E. Rodionov, A. W. Thomas and J. T. Londergan, Mod. Phys. Lett. A9
(1994) 1799.
%
\bibitem{Close88} F.E. Close and A.W. Thomas,
Phys. Lett. B 212 (1988) 227.
%
\end{references}
\end{document}